\documentclass[aps,prl,superscriptaddress,reprint]{revtex4-1}
\usepackage{graphicx}
\usepackage[utf8]{inputenc}
\usepackage{amsmath}
\usepackage{amssymb}
\usepackage{hyperref}
\usepackage{url}
\usepackage{color}
\usepackage{cases}

\newcommand\Eq[1]{Eq.~\ref{#1}}
\newcommand\Fig[1]{Fig.~\ref{#1}}
\newcommand\suppCellSampling{1}
\newcommand\suppPythonDemo{2}
\newcommand\supp[1]{Supp.~#1}
\newcommand\coloronline{\emph{(Color online)}}
\newcommand\diff[1]{\ensuremath{\,\mathrm{d}#1\,}}
\newcommand\ddiff[2]{\ensuremath{\,\mathrm{d}^{#1}#2\,}}
\newcommand\placeholder{{\,\cdot\,}}
\newcommand\dimension{\ensuremath{D}}
\newcommand\bigO[1]{\ensuremath{\OCAL(#1)}}
\newcommand\rmi[2]{\ensuremath{{#1}_{\text{#2}}}}
\newcommand\dirmotion{\ensuremath{\hat{\evec}}}
\newcommand\ptclpair[1]{\ensuremath{\mean{#1}}}
\newcommand\qcell{Q}
\newcommand\qcelltot{\ensuremath{Q_\text{tot}}}
\newcommand\qpart{q}
\newcommand\maxzero[1]{\left[#1\right]^+}
\renewcommand\propto\sim

\let\ifincludesupplements\iftrue
\let\ifnotbuildingseparatesupp\iftrue

\newcommand{\eq}[1]{Eq.~(\ref{#1})}
\newcommand{\eqtwo}[2]{Eqs~(\ref{#1}) and~(\ref{#2})}
\newcommand{\fig}[1]{Fig.~\ref{#1}}
\newcommand{\quot}[1]{``#1''}

\newcommand{\CCAL}{\mathcal{C}}  
\newcommand{\OCAL}{\mathcal{O}}  
\newcommand{\expb}[1]{\exp \glb #1 \grb} 
\newcommand{\expc}[1]{\exp \glc #1 \grc} 
\newcommand{\glb}{\left(}  
\newcommand{\grb}{\right)}  
\newcommand{\glc}{\left[}  
\newcommand{\grc}{\right]}  

\newcommand{\VEC}[1]{\mathbf{#1}}
\newcommand{\cvec}{\VEC{c}}

\newcommand{\evec}{\VEC{e}}

\newcommand{\rvec}{\VEC{r}}

\newcommand{\xvec}{\VEC{x}}

\newcommand{\deltavec}{\boldsymbol{\delta}}

\newcommand{\mean}[1]{\left\langle #1 \right\rangle}

\date{\today}
\begin{document}

\title{Cell-veto Monte Carlo algorithm for long-range systems}

\author{Sebastian C. Kapfer}
\email{sebastian.kapfer@fau.de}
\affiliation{Theoretische Physik 1, FAU Erlangen-Nürnberg, Staudtstr.\ 7, 91058
Erlangen, Germany}

\author{Werner Krauth}
\email{werner.krauth@ens.fr}
\affiliation{Laboratoire de Physique Statistique, Ecole Normale Sup\'{e}rieure
/ PSL Research University, UPMC, Universit\'{e} Paris Diderot, CNRS, 24 rue
Lhomond, 75005 Paris, France}

\ifnotbuildingseparatesupp
\begin{abstract}
We present a rigorous efficient event-chain Monte Carlo algorithm for 
long-range interacting particle systems.
Using a cell-veto scheme within the factorized Metropolis 
algorithm, we compute each single-particle move  with a fixed
number of operations. For slowly decaying potentials
such as Coulomb interactions, screening line charges allow us 
to take into account periodic boundary conditions.
We discuss the performance of the cell-veto Monte 
Carlo algorithm for general inverse-power-law potentials, and illustrate how it 
provides a new outlook on one of the prominent bottlenecks in large-scale 
atomistic Monte Carlo simulations.
\end{abstract}
\maketitle 
\fi

Markov-chain Monte Carlo is one of the most widely used
computational methods in the natural sciences. It samples a high-dimensional 
space of configurations $c$ according to a probability distribution $\pi(c)$.
In the physical sciences, $\pi$ generally corresponds to the Boltzmann 
distribution $\pi(c) = \expc{-\beta E(c)}$, where $\beta$ is the 
inverse temperature and $E$ the system energy.  
The core of most Monte Carlo computations is the Metropolis algorithm 
\cite{Metropolis1953},
which accepts a trial move from configuration $i$ to configuration $f$ with 
probability 
\begin{equation}
p^{\text{Met}}(i \to f) = \min\left\{ 1, \expc{-\beta (E(f) - E(i))} 
\right\}.
\label{eqMetropolis}
\end{equation}
The acceptance probability \Eq{eqMetropolis} satisfies the detailed balance condition,
$\pi(i) p^{\text{Met}}(i \to f) = \pi(f) p^{\text{Met}}(f \to i)$, that 
leads to exponential convergence towards the stationary distribution $\pi(c)$, 
if ergodicity is assured \cite{SMAC}. Moving from one configuration to another 
requires evaluating the induced change of the system energy. In most classical 
$N$-particle simulations, the system energy is a sum over pair terms: $E = 
\sum_{\ptclpair{k,l}} U_{kl} = \sum_{\ptclpair{k,l}} U(\rvec _{kl})$ with the 
pair potential $U$ and the interparticle distances $\rvec_{kl}=\rvec 
_l-\rvec_k$. The evaluation of the system energy generally takes $\bigO{N^2}$ 
operations, and the computation of the energy change upon moving a single 
particle takes $\bigO{N}$  operations.
For a potential with finite support, the change of the system energy for 
moving one particle is computed in \bigO{1}.
To speed up the evaluation, potentials with infinite support, such as the 
Lennard-Jones and other moderately long-ranged potentials, are truncated beyond 
an effective interaction range. This approximation is however known to alter 
the equilibrium properties \cite{Smit1992,intVeldGrest2007}. Strongly 
long-ranged potentials, as 
they appear in electrostatics and gravity, do not allow for the definition of a 
finite interaction range 
and require specialized techniques for determining the system energy to high 
precision.  
Ewald summation \cite{deLeeuw1980,frenkelsmit2001}, for example, adds and subtracts smooth 
charge distributions localized around the point particles.  With periodic 
boundary conditions, this turns the long-ranged part of the interaction into a 
rapidly converging sum in Fourier space. Ewald summation computes the system 
energy in $\bigO{N^{3/2}}$, taking into account periodically replicated images 
of the particles \cite{Perram1988,frenkelsmit2001}. Its refinements further reduce the 
burden of the system-energy computation by discretizing the charge density 
\cite{EastwoodHockneyPPPM} or by exploiting large-scale 
uniformity \cite{FastMultipole}.  Still, in many outstanding applications in the 
natural sciences, the evaluation of long-ranged potentials remains a 
computational bottleneck.  Implementing Ewald summation is particularly 
difficult if periodic boundary conditions are not realized in all dimensions, as 
for example in slab geometries \cite{Brodka2004,MazarsEwald2011}.

In this paper, we present a rigorous Monte Carlo algorithm for $NVT$ particle 
systems with long-ranged interactions that does not evaluate the 
system energy, in contrast to virtually all existing Markov-chain Monte Carlo 
algorithms \cite{SMAC}. This change of perspective opens up many opportunities:
Based on a cell-veto scheme within the factorized Metropolis 
algorithm \cite{MichelKapferKrauth2013}, it 
implements a single-particle move in complexity \bigO{1} without any truncation 
error. For moderately long-ranged potentials, such as Lennard-Jones or dipolar 
interactions, the step size is independent of the system size, and the 
algorithm is effectively constant-time. For strongly long-ranged interactions, 
as the Coulomb forces, the single-move step size slightly 
decreases with $N$. For concreteness, we will consider a fixed 
hypercubic box of 
size $L^\dimension$ with periodic boundary conditions, where \dimension\ is the 
dimension of physical space. The generalization to slab geometries is 
straightforward.

\begin{figure}[htbp]
  \includegraphics[width=\linewidth]{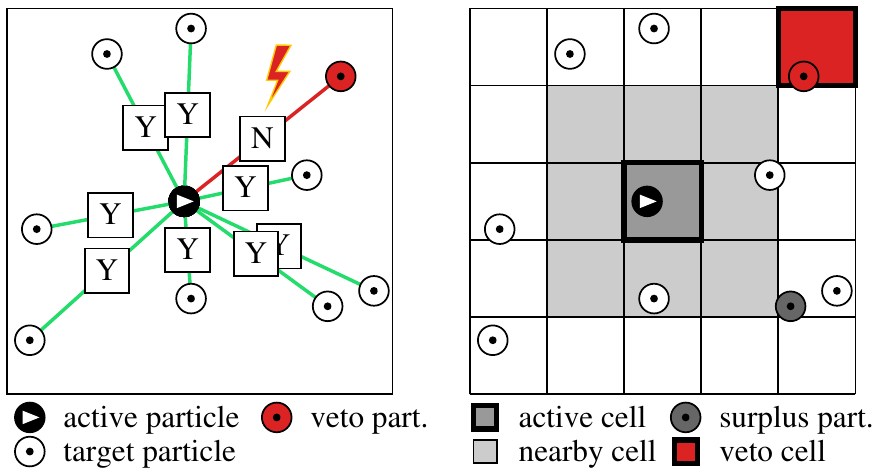}%
  \caption{\coloronline\ Pairwise factorized Monte Carlo 
algorithm. \emph{Left:} The move of the active particle is vetoed by one target 
particle, so that the necessary consensus (all \quot{Y})
is not reached (see \eq{eqFactMetropolis}). \emph{Right:} In the 
cell-veto algorithm, vetos are provisionally solicited on the cell level
(between the active cell $\CCAL_a$ and the target cell $\CCAL_t$) before being 
confirmed for the active particle, at $\rvec_a \in \CCAL_a$ and the target 
particle, at $\rvec_t \in \CCAL_t$. Nearby and surplus particles are treated 
differently.}
   \label{cell_schema}
\end{figure}

In contrast to the Metropolis algorithm of \Eq{eqMetropolis}, the 
pairwise factorized algorithm\cite{MichelKapferKrauth2013} accepts moves with 
the probability
\begin{equation}
    p^{\text{fact}}(i \to f)
=  
    \prod_{\ptclpair{k,l}}\underbrace{\min\left\{ 1, \expc{-\beta \Delta U_{kl}(i\to f)
        } \right\}}_{p_{kl}(i \to f)},
\label{eqFactMetropolis}
\end{equation}
where $\Delta U_{kl}$ is the change in the pair potential between particles $k$ 
and $l$.  In our algorithm, we never explicitly evaluate the
function $p^{\text{fact}}$. Rather, the product of probabilities on 
the rhs of \eq{eqFactMetropolis} is interpreted as a condition that is true if 
all its factors are true. The move $i \to f$ is thus accepted by consensus, 
namely if each pair $\ptclpair{k,l}$ independently accepts the move with 
probability $p_{kl}$ \cite{MichelKapferKrauth2013} (see \fig{cell_schema}). 
Instead of computing the energy to high precision, we will compute upper bounds 
for the veto probability $1 - p_{kl}$ by embedding particles $k$ and $l$ into 
cells $\CCAL_k$ and $\CCAL_l$, respectively. To identify particles 
vetoing
the move, one rapidly identifies cell vetos and inspects the contents of 
corresponding cells to determine whether the cell vetos are confirmed on the 
particle level (see \fig{cell_schema}).

In continuum space, two configurations $i$ and $f$ with $i \ne f$ can be 
infinitesimally close to each other. For regular potentials, this implies that 
the change of pair energies $\Delta U_{kl}(i\to f)$, and therefore the veto 
probability $1 - p_{kl}$, are infinitesimal as well. In the event-chain 
algorithm \cite{Bernard2009,MichelKapferKrauth2013}, a proposed move $i \to 
f$ consists in 
the infinitesimal displacement of an \quot{active} particle $a$ in a direction 
\dirmotion: The proposed move is 
$ \rvec_a(i) \to \rvec_a(f) = \rvec_a(i) + \dirmotion \diff s$, where $\diff s$ 
is an 
infinitesimal time increment. The active particle keeps moving in the same 
direction until a move is finally vetoed by a target particle $t$. The target 
particle then becomes the new active particle, i.\,e., the proposed move is $(i, 
a, \dirmotion) \to (f, a, \dirmotion)$, and if vetoed by particle pair 
$\ptclpair{a,t}$, the configuration is changed to $(i, t, \dirmotion)$. 
This implements a \quot{lifted} Markov chain \cite{Diaconis2000} 
with two additional variables $a$ and $\dirmotion$, which trivially projects to 
the physical space with the proper Boltzmann distribution. Veto probabilities $1-p_{at}$ are infinitesimal. 
Two simultaneous vetos are thus prevented from arising from different target 
particles. 
Detailed balance is violated (the reverse move $\rvec_a(f) = \rvec_a(i) - 
\dirmotion\diff s$ is never proposed). However, the event-chain algorithm 
satisfies the global-balance condition
\begin{equation}
    \sum_i \pi(i) p( i \to f) = \pi(f)
\label{eqGlobalBalance}
\end{equation}
sufficient for exponential convergence to the equilibrium distribution on the 
accessible configurations.  To ensure ergodicity, both the active particle and 
the direction of motion are periodically reset to random values (see 
\supp\suppPythonDemo). Lifted Markov chains have been shown to improve 
convergence speed in many cases, and also to lower the dynamical critical 
scaling exponents \cite{Diaconis2000,KapferKrauth2015,MichelMayerKrauth2015,Nishikawa2015}.

The core of an event-chain program consists in determining the step size
$\Delta s$ to the next particle event and in identifying the vetoing target
particle $t$, rather than explicitly programming small time increments
(see \Fig{figDisplace}a).  The actual move then merely consists in updating the
active particle position as $\rvec_a \to \rvec_a + \dirmotion\Delta s$ and in
changing the active particle to $t$.
For long-ranged potentials, $t$ can be far away from the active particle.
At any instant during the simulation, the veto 
probability of a potential target particle $t$ is given by the particle-event rate $\qpart$, 
defined via a directional derivative of the pair potential,
\begin{align}
     1-p_{at} = \qpart(\rvec_{at}) \diff s = \beta 
      \maxzero{-\dirmotion\cdot\boldsymbol\nabla U_{at}} \diff s
\end{align}
with $\maxzero{\placeholder} = \max(0, \placeholder)$.
For long-ranged potentials, $\qpart$ carries over large distances (see 
\Fig{figDisplace}b, c).
Particle-event distances $r_{at}$ are distributed as $\qpart(r_{at}) g(r_{at})$,
where $g$ is the radial distribution function, and thus exhibit the same long-ranged tail.
In contrast, the displacement between events, i.\,e., the step size $\Delta s$,
decays exponentially within a few interparticle distances, see \fig{figDisplace}c.
For each pair $\ptclpair{a,t}$, the event 
time $\Delta s_t$  can be computed in \bigO{1}, so that the event-chain 
algorithm can be implemented in $\bigO{N}$ per particle event \cite{Peters2012}, 
by iterating over all target particles.  The earliest veto will define the
step size $\Delta s$ and the active particle for the next step.

\begin{figure}[b]
  \includegraphics{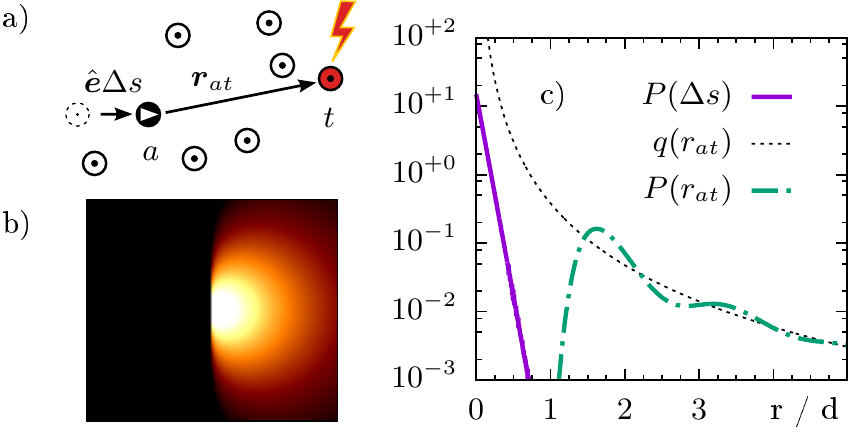}%
  \caption{\coloronline~
  Event-chain algorithm for a long-ranged dipolar potential in two dimensions,
  $\beta U = 10 \times (d/r)^3$.
  \emph{a)} The active particle takes infinitesimal moves in the $+x$ direction.
  At time $\Delta s$, a move is vetoed by particle $t$, at event-time
  distance $\rvec_{at}$.  The vetoing particle $t$ becomes the new active 
  particle and starts  to move in the $+x$ direction.
  \emph{b)} Heatmap representation of the particle-event rate 
$\qpart(\rvec_{at})$.
  The active particle is in the center, black corresponds to $\qpart = 0$.
  \emph{c)} Probability distributions of the step size $\Delta s$ taken by the
  active particle and of particle-event distances $r_{at}$. }
\label{figDisplace}
\end{figure}

For a homogeneous system (with a bounded particle density), the complexity per 
particle event can be reduced from \bigO{N} to \bigO{1} by establishing 
upper bounds for the particle-event rate which hold irrespective of the 
precise particle positions. Concretely, we superimpose a fixed regular grid 
onto the system, with cells typically containing at most one particle (see 
\fig{cell_schema}; rare 
\quot{surplus} particles are treated separately). The 
particle-event rate between the active particle in cell $ \CCAL_a$ and a 
target particle in cell $\CCAL_t$ is bounded from above by the 
cell-veto rate 
\begin{equation}
\qcell ( \CCAL_a, \CCAL_t) = \max_{\rvec_a \in \CCAL_a, \rvec_t \in \CCAL_t} 
    \qpart(\rvec_t-\rvec_a).
\label{eqCellSelectRate}
\end{equation}
This quantity depends only on the pair potential and the relative positions of 
the two cells and can be tabulated before sampling starts.  The cell-veto rate 
remains finite except for a few nearby cells that contain the hard-core
singularities.  In the case of point particles, these must include any cells that share
corners with $\CCAL_a$ (see \fig{cell_schema}).  For efficiency,
\quot{nearby} cells may comprise a larger portion of the short-range 
features of $U$.

Excluding nearby and surplus particles, the total particle-event rate is
bounded from above by the total cell-veto rate
\begin{equation}
 \qcelltot = \sum_{\CCAL_t} \qcell( \CCAL_a, \CCAL_t),
 \label{e:total_cell_event_rate}
\end{equation}
which remains a constant throughout the simulation. The next cell veto can 
then be sampled in \bigO{1}: The time is distributed exponentially
\begin{equation}
    P(\Delta s) = \qcelltot \expb{- \qcelltot \Delta s},
\end{equation}
so that $\Delta s$ is given through the logarithm of a uniform random number 
(\cite{SMAC}, see \supp\suppPythonDemo). The cell veto is triggered by the cell $\CCAL_t$
with probability $\propto \qcell(\CCAL_{a}, \CCAL_{t})$.  The selection of the target
cell from all the non-nearby cells can also be accomplished in constant time (see below).
If the vetoing cell $\CCAL_t$ contains a particle, at position $\rvec_t$, it is then 
chosen as the target particle for a particle event with probability 
$\qpart(\rvec_a + \dirmotion\Delta s,\rvec_t) / \qcell(\CCAL_{a}, \CCAL_{t})$.
This long-range particle event must be put into competition with events
triggered by nearby or surplus particles, which are handled as in the
short-range event-chain algorithm \cite{MichelKapferKrauth2013} (see also
\supp\suppPythonDemo).  The number of
nearby particles is naturally bounded. The number of surplus particles may be
kept as small as desired by adapting the cell size. In practice, we use cells
that are sufficiently small so that surplus particles appear only
exceptionally. Consequently, a cell veto can effectively be processed constant
time, and the performance of the cell-veto algorithm depends on the rate of
cell vetos \qcelltot.

The total cell-veto rate $\qcelltot$ depends on the range of the pair potential. 
For inverse-power-law interactions, $U(r) \propto 1/r^n$, the event rate for a 
bare particle
scales as $\qpart \propto 1/r^{n+ 1}$ \footnote{We set, for $n\not=0$,
$\beta U(r) =\beta\varepsilon\times ((d/r)^n - 1)/n$ and
for $n=0$, $\beta U(r) = \beta\varepsilon\times  \ln(d/r)$, i.~e., planar 
Coulomb;
here $\varepsilon$ and $d$ are the scales of energy and length.
In either case, $\tilde q(\rvec) = \beta\varepsilon \times 
(\dirmotion\cdot\rvec) / r^{2+n}$.
}.
In an infinite system, the total cell-veto rate $\qcelltot \propto 
\int\!\ddiff{D}{r} \qpart$ is finite for moderately long-ranged potentials, 
i.~e., for $n > D-1$. This class includes dipolar forces in $D=2$ and $D=3$, as 
well as the Lennard-Jones potential. In this case, the cell-veto algorithm is of 
complexity $\bigO{1}$.

\begin{figure}[htbp]
    \centering%
    \includegraphics{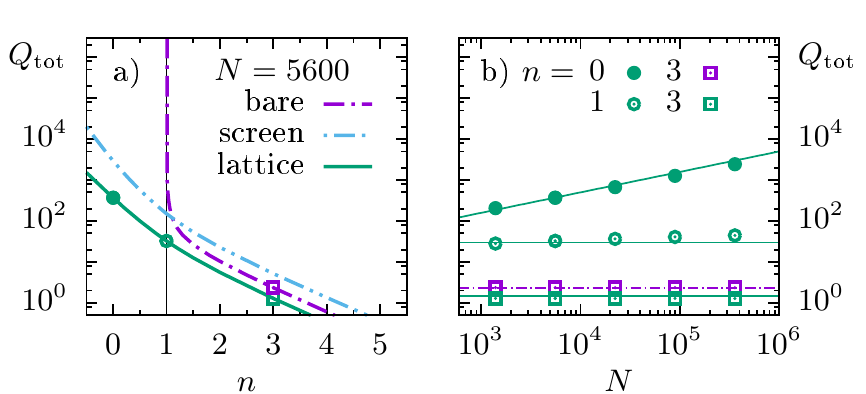}%
    \caption{\coloronline~%
\emph{a)} Total cell-veto rate \qcelltot, \Eq{eqCellSelectRate}, for inverse-power-law potentials in $D=2$.
The event rate for bare particles diverges as $n\to D-1$ (vertical line) because
of the presence of periodic images, while screened event rates stay finite.
Solid bullets are the Coulomb system ($n=D-2$; for 2D, $U(r)\propto -\ln r$).
\emph{b)} Scaling of \qcelltot\ with system size $N$, for the screened-lattice
algorithm (green solid), and for the bare-particle algorithm (purple dashed).
The inclined line is $\propto N^{1/2}$.
}
    \label{figCellSelectRate}
    \label{figQtotDivg}
\end{figure}

For strongly long-ranged potentials ($n \le D-1$,
including Coulomb forces), the cell-veto rate in an infinite system diverges
(see \Fig{figQtotDivg}a).  In the replicated-box representation of periodic
boundary conditions (see \fig{figLineCharges}), even the sum over all periodic
images of a single target particle ($N=2$) leads to an infinite particle-event
rate.
The sum may be regularized by adding uniformly charged line segments (parallel to the
direction of motion \dirmotion) that neutralize each particle charge yet combined
leave invariant the energy differences of the original system.  Screening line
charges can be defined for general potentials. For inverse-power-law
interactions, the directional derivatives of the particle and line-charge
potentials are
\begin{align}
\tilde{q}(\rvec) & = \beta\varepsilon d^n \times \frac{\dirmotion \cdot \rvec} 
{r^{n+2}}\label{dir_derivative},\\
\tilde{l}(\rvec) & = \frac{\beta}{L} \times
\Bigl[ U(\rvec+ \dirmotion L / 2) - U(\rvec- \dirmotion L / 2) \Bigr], \label{line_derivative}
\end{align}
where $\rvec$ is the folded-out distance vector between the active particle and 
a particular periodic image of the target particle.
By vanishing monopole and dipole moments, $\tilde q+\tilde l$ asymptotically
decays as $1/r^{n+3}$, sufficient to render $\qcelltot$ unconditionally
convergent for Coulomb forces.

We may now define three distinct particle-event rates:
\begin{numcases}
    {\qpart(\rvec)  = }
\maxzero{\tilde{q}(\rvec)} & \text{bare}, \label{bare_q}\\
\maxzero{\tilde{q}(\rvec) + \tilde{l}(\rvec)} & \text{screened}, 
\label{screened_q}\\
\Bigl[ \sum_{k, \text{p.i.}}\tilde{q}(\rvec_k) + 
\tilde{l}(\rvec_k)\Bigr]^+\! \! \! \! \! \! \! \! \! \! \! & \text{screened lattice}. 
\label{screened_lattice_q}
\end{numcases}
The screened-lattice version of \eq{screened_lattice_q}, where the sum extends
over all periodic images of the target particle, minimizes the cell-veto rate
by merging the periodic images into the primary copy of each particle.  The
number of target cells $\CCAL_t$ is finite, and the target cell of a cell veto
can be found extremely efficiently by precomputing the function $\qcell(\CCAL_a, \CCAL_t)$ and
employing Walker's alias method or related techniques
\cite{Walker:1977,Marsaglia2004} (see \supp\suppCellSampling).
A commented Python implementation of the cell-veto Monte Carlo algorithm using
this approach is provided in \supp\suppPythonDemo.

In an alternative version of the cell-veto algorithm, the particle-event rates
of \eq{bare_q} and \eq{screened_q} are used with explicitly replicated
simulation boxes.  An infinite number of target cells are considered.  The
target cell for a cell veto can still be found in constant time by rejection
sampling.  A vector $\rvec$ is sampled with probability density $\propto
Q(\rvec)$, where $Q$ is an upper bound to the particle-event rate $Q(\rvec)\geq
q(\rvec+\deltavec)$ for all vectors $\deltavec$ shorter than the cell diagonal.
The target cell $\CCAL_t$ is then the cell containing the point $\rvec_a + \rvec$
(see \supp\suppCellSampling).
The cell-veto rates are somewhat larger than for the lattice-screened version.
This may however be offset by the less onerous evaluation of \eq{bare_q} or
\eq{screened_q} compared to \eq{screened_lattice_q} (surplus particles must be
treated with the lattice-screened version).

Both the screened and the screened-lattice particle-event rates
overcome the divergence at $n = D-1$ with periodic boundary conditions (see
\Fig{figQtotDivg}a). Since one cell veto can be handled in $\bigO{1}$ operations,
the computational cost of simulating a fixed timespan is proportional to
the rate of cell vetos.
For a distance vector $\rvec = L\cvec$, the directional derivatives in 
\eqtwo{dir_derivative}{line_derivative} scale as $\propto L^{-n-1}$ and so do 
the particle-event rates $\qpart$. This implies that, above the point $n=D-1$,
the cell-veto rate at constant density scales as \bigO{L^{\dimension-n-1}} =
\bigO{N^{1 - (n+1)/\dimension}}.  For Coulomb forces in $D$ dimensions, $n=D-2$,
we find $\qcelltot\sim N^{1/D}$, see \Fig{figCellSelectRate}b.  Thus, in three
dimensions, the cell-veto algorithm is of complexity~\bigO{N^{1/3}}.  This 
compares
favorably with the cost of an \bigO{N^{3/2}} energy evaluation with Ewald
summation in conventional Metropolis Monte Carlo.

\begin{figure}[htbp]
    \centering%
\includegraphics[width=0.8\columnwidth]{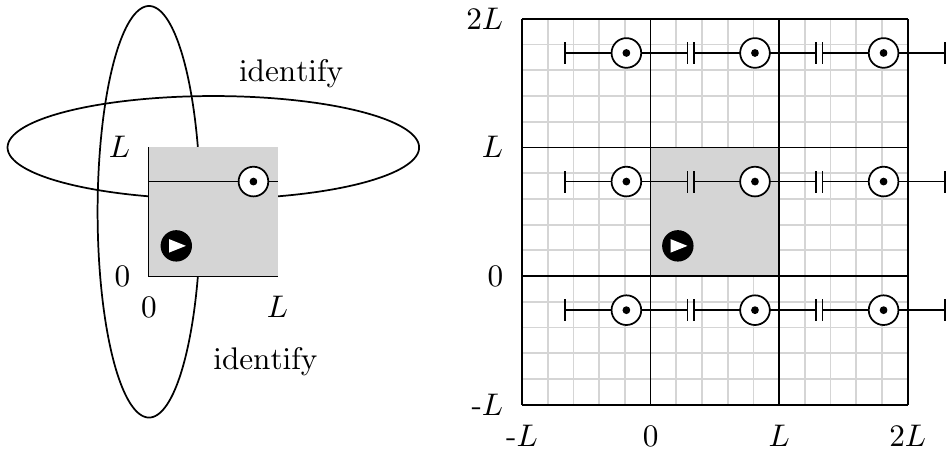}
    \caption{ Periodic particle system with screening line charges.
    \emph{Left:} Active and target particles and screening line charge segments
in a periodic two-dimensional square box. \emph{Right:} Folded-out periodic 
system with image target particles, each of which forms a
neutral composite particle together with its screening charge.}
    \label{figLineCharges}
\end{figure}

In conclusion, we have presented a cell-veto Monte Carlo algorithm that need not
compute the system energy. Remarkably, it advances the 
physical state of the system by one event in \bigO{1} even for long-ranged 
interactions. The algorithm introduces none of the cutoffs that come with 
practical versions of Ewald summation. Strongly long-ranged potentials such as 
electrostatic forces are handled exactly using screening line charges. The 
complexity of the algorithm then scales weakly with $N$. It is hoped that the 
algorithm will permit to access much larger systems than was previously 
possible. The demo program of  \supp\suppPythonDemo, and the 
C++ version of this algorithm are available online \cite{PostLHC}.

\bibliography{literature}

\ifincludesupplements

\clearpage

\title{Supplement to ``Cell-veto Monte Carlo algorithm for long-range systems''}

\maketitle

\onecolumngrid
\renewcommand{\theequation}{S\arabic{equation}}
\renewcommand{\thefigure}{S\arabic{figure}}
\setcounter{equation}{0}
\setcounter{figure}{0}
\setcounter{page}{1}

\subsection{Supplementary Item \suppCellSampling: Cell-veto sampling}

The pairwise factorized Metropolis algorithm determines pair events (vetos) and 
thus avoids to compute the system energy. The cell-veto algorithm takes this 
strategy one step farther. Instead of scanning all particle pairs for vetos, it 
first solicits cell vetos (see \fig{cell_schema}), which then have to be 
confirmed on the level of the actual particle positions. Even with periodic boundary 
conditions, the number of cells remains finite if all the periodic images of a 
particle are merged into the one located in the primary simulation box (see
\eq{screened_lattice_q}).  The next cell veto must be selected from the $\rmi{N}{cell}$
cells $\CCAL_t$ with a nonzero cell-veto rate.  Each cell must be sampled
with probability $\propto\qcell(\CCAL_a, \CCAL_t)$, see \fig{walker_schema}.
This finite discrete-probability sampling problem is best solved through a
rejection-free exact algorithm, as Walker's alias method.  In Walker's method,
the cell-veto rates are reassembled into composite rates consisting of at most
two original rates and adding up to exactly the mean cell-veto rate
$Q_\text{mean} = \qcelltot/ \rmi{N}{cell}$. The cutting-up and reassembling of the $\qcell(\CCAL_a,
\CCAL_t)$ constitutes the initialization stage
of Walker's method (in the demo program of \supp\suppPythonDemo: 
in function \verb!WalkerSet!).  In the sampling stage, a cell $\CCAL_t$ can be sampled
with the proper probability by first sampling the composite rate (as a random integer between
$1$ and $\rmi{N}{cell}$) and then deciding between 
the at most two rates by sampling a uniform random real between $0$ and 
$Q_\text{mean}$ (\verb!WalkerSample!, in the demo program of \supp\suppPythonDemo).
This step is constant time and independent of the number of cells.

\begin{figure}[htbp]
  \includegraphics[width=0.7\linewidth]{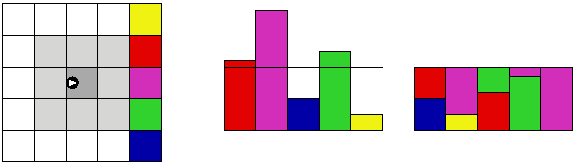}
  \caption{Cell-veto sampling using Walker's method.
  \emph{Left:} The non-nearby cells (5 such cells shown in different 
   colors) may all have finite cell-veto rates.
  \emph{Center:} Cell-veto rates in a linear 
  representation. The mean cell rate $Q_\text{mean}$ is indicated (5 cells 
  shown, again).
  \emph{Right:} In the initialization stage of Walker's method, the cell 
  rates are reassembled, at most by pairs, into composite rates. }
   \label{walker_schema}
\end{figure}

\begin{figure}[htbp]
  
\includegraphics[width=0.6\linewidth]{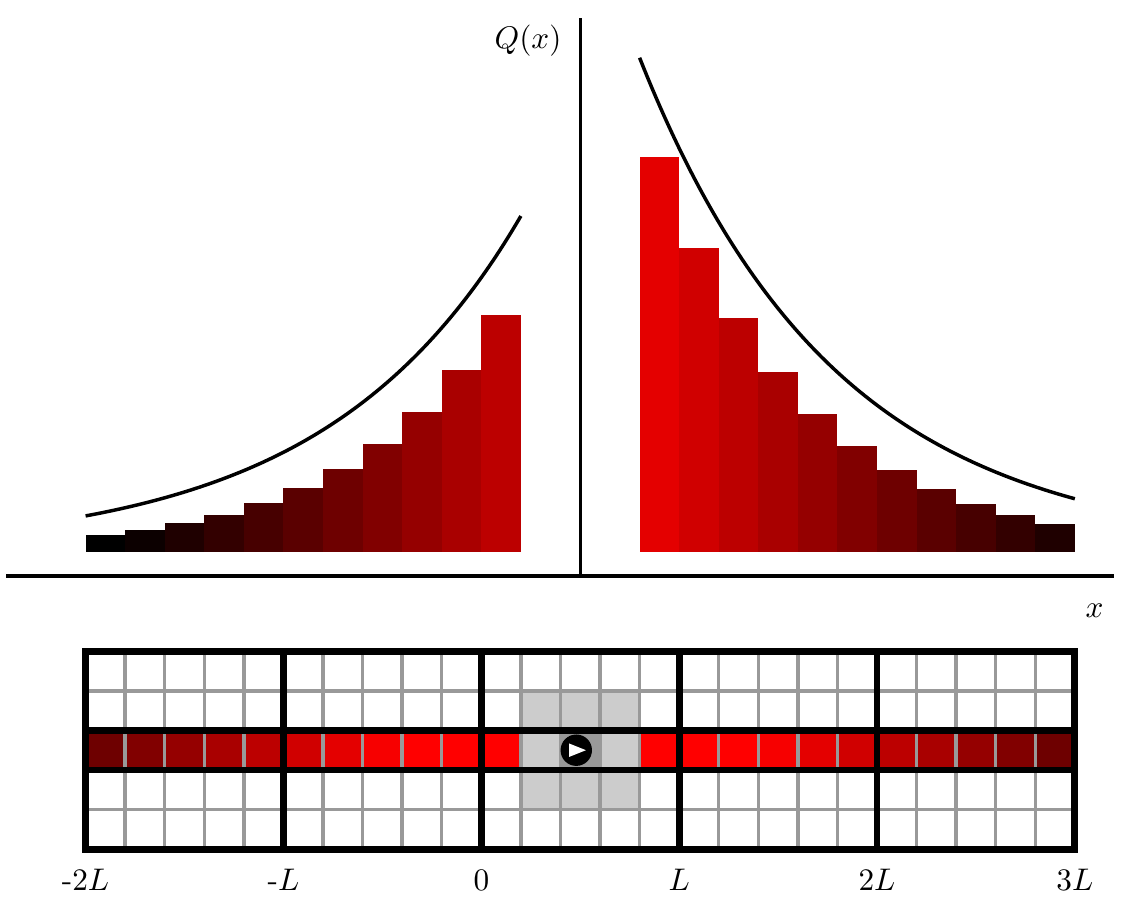}
  \caption{Cell-veto sampling using rejection sampling with a function $Q(x)$
 in a \quot{folded-out} periodic system and an infinite number of 
cells (one-dimensional representation shown). 
  }
  \label{unfolded_rejection}
\end{figure}

Alternatively, one may also keep the individual cells, that is, work 
explicitly in the folded-out version of the system, and with the cell-veto rates of 
\eq{screened_q} that consider each periodic copy of a target cell individually.
The number of cells is now countably infinite. Nevertheless, it is easy to 
devise a rejection-sampling strategy using a function that is easy to sample, integrable to infinity,
and an upper bound to the cell-veto rate (see \fig{unfolded_rejection} 
for a one-dimensional representation).  A point $x$ sampled from the probability
distribution $\propto Q(x)$ identifies a cell.  If that cell contains a target particle $t$,
a particle event is triggered with probability $q(\rvec_t) / \rmi{V}{cell} / Q(x)$.
In the folded-out formulation of the cell-veto algorithm, surplus particles must 
still be merged with their periodic images, in order to keep their number finite. 

\subsection{Supplementary Item \suppPythonDemo: Demo implementation of the cell-veto
algorithm}
The demo implementation of the cell-veto Monte Carlo algorithm, 
the program \verb!demo_cell.py!, is written in the Python~2 programming 
language. 
$N$ particles are simulated in a two-dimensional square box of length $1$ with 
periodic boundary conditions, and with an $1/r$ pair potential that is 
periodically continued. A regular square grid with $L^2$ cells is superimposed 
to the system. Cells are numbered from $0$ to $L^2-1$. The screened-lattice 
particle-event rate of \eq{screened_lattice_q} is implemented (naively).
Walker's method is used for sampling the veto cells.

In the setup stage of \verb!demo_cell.py!, particles
are initialized to  random positions, and the cell-veto rates are computed 
between the active cell 
$\CCAL_a = 0$ and all other target cells that are not nearby $\CCAL_a = 0$. 
The function \verb!translated_cell! transfers this calculation (with $\CCAL_a 
=0$)  to arbitrary cell pairs  $(\CCAL_a, \CCAL_t)$.
Specifically, the cell-veto rate is defined as the maximum of the
particle-event rate over all positions, as indicated in 
\eq{eqCellSelectRate}. For this demo program, it is assumed that the maximum 
particle-event 
rate is attained for $\xvec_a $ and $\xvec_t$ on the boundary of $\CCAL_a$ and 
$\CCAL_t$, respectively, and discrete points in the list \verb!cell_boundary! are 
used. For the demo version, the lattice-screened particle-event rate of 
\eq{screened_lattice_q} is determined by a naive direct summation of the images 
of the target particle and its screening line charge (see function 
\verb!pair_event_rate!), 
rather than by an efficient function evaluation. The initializaton of Walker's 
alias method, as explained in \supp\suppCellSampling, concludes the 
setup stage of \verb!demo_cell.py!.

In one iteration of the sampling stage of \verb!demo_cell.py!, particles 
advance by a total distance \verb!chain_ell! (see \cite{Bernard2009,MichelKapferKrauth2013}) in a fixed direction. 
This direction of motion is first sampled (from $+x$ or $+y$). In the demo 
version, only the $+x$ move is implemented explicitly ($+y$ moves are 
implemented indirectly by flipping all particle coordinates $(x_i, y_i) \to 
(y_i, x_i)$).
At the beginning of this iteration (given that such a flip may have taken 
place)  particles are reclassified into target 
particles associated to cells (at most one per cell), and surplus particles.
(Each cell must contain at most one particle, in order for the 
cell-veto rate to be an upper limit for the particle-event rate from all 
particles within the cell). The 
active particle is then sampled uniformly among all particles in the system.  
At each step of the iteration, the step size \verb!delta_s! to the next cell veto 
is sampled from the total cell-veto rate $\qcelltot$.
The cell veto may be preempted by the end of the chain, after displacement \verb!chain_ell!.
It is also checked whether the cell veto occurs after the active particle crosses
the cell limit: We must trigger an event when the cell 
boundary is reached, as the set of nearby particles then changes.
If the cell veto is indeed confirmed on the particle level, it is 
put into competition with events triggered by nearby or surplus particles.
In the demo version, the particle-event rates for nearby or surplus particles 
are computed in a simplified way.

The \verb!demo_cell.py! program (see below) was tested against a straightforward
implementation of the Metropolis algorithm, and against the C++ version (see
\url{https://www.github.com/cell-veto/postlhc/}).

\begin{verbatim}
import math, random, sys
import numpy as np

def norm (x, y):
    """norm of a two-dimensional vector"""
    return (x*x + y*y) ** 0.5

def dist (a, b):
   """periodic distance between two two-dimensional points a and b"""
   delta_x = (a[0] - b[0] + 2.5) % 1.0 - 0.5
   delta_y = (a[1] - b[1] + 2.5) % 1.0 - 0.5
   return norm (delta_x, delta_y)

def random_exponential (rate):
    """sample an exponential random number with given rate parameter"""
    return -math.log (random.uniform (0.0, 1.0)) / rate

def pair_event_rate (delta_x, delta_y):
    """compute the particle event rate for the 1/r potential in 2D (lattice-screened version)"""
    q = 0.0
    for ky in range (-k_max, k_max + 1):
        for kx in range (-k_max, k_max + 1):
            q += (delta_x + kx) / norm (delta_x + kx, delta_y + ky) ** 3
        q += 1.0 / norm (delta_x + kx + 0.5, delta_y + ky)
        q -= 1.0 / norm (delta_x - kx - 0.5, delta_y + ky)
    return max (0.0, q)

def translated_cell (target_cell, active_cell):
    """translate target_cell with respect to active_cell"""
    kt_y = target_cell // L
    kt_x = target_cell  % L
    ka_y = active_cell // L
    ka_x = active_cell  % L
    del_x = (kt_x + ka_x) % L
    del_y = (kt_y + ka_y) % L
    return del_x + L*del_y

def cell_containing (a):
    """return the index of the cell which contains the point a"""
    k_x = int (a[0] * L)
    k_y = int (a[1] * L)
    return k_x + L*k_y

def walker_setup (pi):
    """compute the lookup table for Walker's algorithm"""
    N_walker = len(pi)
    walker_mean = sum(a[0] for a in pi) / float(N_walker)
    long_s = []
    short_s = []
    for p in pi:
        if p[0] > walker_mean:
            long_s.append (p[:])
        else:
            short_s.append (p[:])
    walker_table = []
    for k in range(N_walker - 1):
        e_plus = long_s.pop()
        e_minus = short_s.pop()
        walker_table.append((e_minus[0], e_minus[1], e_plus[1]))
        e_plus[0] = e_plus[0] - (walker_mean - e_minus[0])
        if e_plus[0] < walker_mean:
            short_s.append(e_plus)
        else:
            long_s.append(e_plus)
    if long_s != []: 
        walker_table.append((long_s[0][0], long_s[0][1], long_s[0][1]))
    else: 
        walker_table.append((short_s[0][0], short_s[0][1], short_s[0][1]))
    return N_walker, walker_mean, walker_table

def sample_cell_veto (active_cell):
    """determine the cell which raised the cell veto"""
    # first sample the distance vector using Walker's algorithm
    i = random.randint (0, N_walker - 1)
    Upsilon = random.uniform (0.0, walker_mean)
    if Upsilon < walker_table[i][0]:
        veto_offset = walker_table[i][1]
    else:
        veto_offset = walker_table[i][2]
    # translate with respect to active cell
    veto_rate = Q_cell[veto_offset][0]
    vetoing_cell = translated_cell (veto_offset, active_cell)
    return vetoing_cell, veto_rate


N = 40
k_max = 3 # extension of periodic images.
chain_ell = 0.18  # displacement during one chain
L = 10  # number of cells along each dimension
density = N / 1.
cell_side = 1.0 / L

# precompute the cell-veto rates
cell_boundary = []
cb_discret = 10 # going around the boundary of a cell (naive)
for i in range (cb_discret):
    x = i / float (cb_discret)
    cell_boundary += [(x*cell_side, 0.0), (cell_side, x*cell_side),
                      (cell_side - x*cell_side, cell_side),
                      (0.0, cell_side - x*cell_side)]

excluded_cells = [ del_x + L*del_y for del_x in (0, 1, L-1) \
                                   for del_y in (0, 1, L-1) ]
Q_cell = []

for del_y in xrange (L):
    for del_x in xrange (L):
        k = del_x + L*del_y
        Q = 0.0
        # "nearby" cells have no cell vetos
        if k not in excluded_cells:
            # scan the cell boundaries of both active and target cells
            # to find the maximum of event rate
            for delta_a in cell_boundary:
                for delta_t in cell_boundary:
                    delta_x = del_x*cell_side + delta_t[0] - delta_a[0]
                    delta_y = del_y*cell_side + delta_t[1] - delta_a[1]
                    Q = max (Q, pair_event_rate (delta_x, delta_y))
        Q_cell.append ([Q, k])

Q_tot = sum (a[0] for a in Q_cell)
N_walker, walker_mean, walker_table = walker_setup (Q_cell)

# histogram for computing g(r)
hbins = 50
histo = np.zeros (hbins)
histo_binwid = .5 / hbins
hsamples = 0

# random initial configuration
particles = [ (random.uniform (0.0, 1.0), random.uniform (0.0, 1.0))
    for _ in xrange (N) ]

for iter in xrange (10000):
    if iter % 100 == 0:
        print iter

    # possibly exchange x and y coordinates for ergodicity
    if random.randint(0,1) == 1:
        particles = [ (y,x) for (x,y) in particles ]
    # pick active particle for first move
    active_particle = random.choice (particles)
    particles.remove (active_particle)
    active_cell = cell_containing (active_particle)
    # put particles into cells
    surplus = []
    cell_occupant = [ None ] * L * L
    for part in particles:
        k = cell_containing (part)
        if cell_occupant[k] is None:
            cell_occupant[k] = part
        else:
            surplus.append (part)

    # run one event chain
    distance_to_go = chain_ell
    while distance_to_go > 0.0:
        planned_event_type = 'end-of-chain'
        planned_displacement = distance_to_go
        target_particle = None
        target_cell = None

        active_cell_limit = cell_side * (active_cell % L + 1)
        if active_cell_limit - active_particle[0] <= planned_displacement:
            planned_event_type = 'active-cell-change'
            planned_displacement = active_cell_limit - active_particle[0]

        delta_s = random_exponential (Q_tot)
        while delta_s < planned_displacement:
            vetoing_cell, veto_rate = sample_cell_veto (active_cell)
            part = cell_occupant[vetoing_cell]
            if part is not None:
                Ratio = pair_event_rate (part[0] - active_particle[0] - delta_s, \
                                         part[1] - active_particle[1])           \
                        / veto_rate
                if random.uniform (0.0, 1.0) < Ratio:
                    planned_event_type = 'particle'
                    planned_displacement = delta_s
                    target_particle = part
                    target_cell = vetoing_cell
                    break
            delta_s += random_exponential (Q_tot)

        # compile the list of particles that need separate treatment
        extra_particles = surplus[:]
        for k in excluded_cells: 
            part = cell_occupant[translated_cell (k, active_cell)]
            if part is not None:
                extra_particles.append (part)

        # naive version of the short-range code by discretization
        delta_s = 0.0
        short_range_step = 1e-3
        while delta_s < planned_displacement:
            for possible_target_particle in extra_particles:
                # this supposes a constant event rate over the time interval
                # [delta_s:delta_s+short_range_step]
                q = pair_event_rate (possible_target_particle[0] - active_particle[0] - delta_s,
                                     possible_target_particle[1] - active_particle[1])
                if q > 0.0:
                    event_time = random_exponential (q)
                    if event_time < short_range_step and delta_s + event_time < planned_displacement:
                        planned_event_type = 'particle'
                        planned_displacement = delta_s + event_time
                        target_particle = possible_target_particle
                        target_cell = cell_containing (target_particle)
                        break
            delta_s += short_range_step

        # advance active particle
        distance_to_go -= planned_displacement
        new_x = active_particle[0] + planned_displacement
        active_particle = (new_x % 1.0, active_particle[1])

        if planned_event_type == 'active-cell-change':
            ac_x = (active_cell_limit + 0.5*cell_side) % 1.0
            active_cell = cell_containing ([ac_x, active_particle[1]])
            active_particle = (active_cell % L * cell_side, active_particle[1])

        elif planned_event_type == 'particle':
            # remove newly active particle from store
            if target_particle in surplus:
                surplus.remove (target_particle)
            else:
                cell_occupant[target_cell] = None
            # put the previously active particle in the store
            if cell_occupant[active_cell] is not None:
                surplus.append (active_particle)
            else:
                cell_occupant[active_cell] = active_particle
            active_particle = target_particle
            active_cell = cell_containing (active_particle)

    # restore particles vector for x <-> y transfer
    particles = [ active_particle ]
    particles += [ part for part in cell_occupant if part is not None ]
    particles += surplus

    # form histogram for computing radial distribution function g(r)
    for k in range (len (particles)):
        for l in range (k):
            ibin = int (dist (particles[k], particles[l]) / histo_binwid)
            if ibin < len (histo):
                histo[ibin] += 1
        hsamples += 1

# compute g(r) from histogram
half_bin = .5 * histo_binwid
r = np.arange (0., hbins) * histo_binwid + half_bin
g_of_r = histo / density / hsamples * 2
g_of_r /= math.pi * ((r+half_bin)**2 - (r-half_bin)**2)
# save g(r)
np.savetxt ('cvmc-radial-distr-func.dat', zip (r, g_of_r))
\end{verbatim}

\fi
\end{document}